\documentclass[3p,times,procedia]{elsarticle}
\usepackage{nupha_ecrc}

\volume{00}

\firstpage{1}

\journalname{Nuclear Physics A}

\runauth{Yuanjing Ji}

\jid{nupha}

\jnltitlelogo{Nuclear Physics A}

\usepackage{amssymb}

 \usepackage{lineno}

\usepackage[figuresright]{rotating}

\begin{document}

\begin{frontmatter}



\dochead{XXVIIIth International Conference on Ultrarelativistic Nucleus-Nucleus Collisions\\ (Quark Matter 2019)}

\title{Heavy flavor physics with the sPHENIX MAPS vertex tracker upgrade}


\author[1,2]{Yuanjing Ji (for the sPHENIX Collaboration)}

\address[1]{State Key Laboratory of Particle Detection and Electronics, \\ University of Science and Technology of China, Hefei, Anhui 230026, China}
\address[2]{Lawrence Berkeley National Laboratory, Berkeley, CA 94706, USA}

\begin{abstract}
The sPHENIX detector at the Relativistic Heavy Ion Collider will measure a suite of unique jet and Upsilon observables with unprecedented statistics and kinematic reach. A MAPS-based vertex detector upgrade to sPHENIX, the MVTX, will provide a precise determination of the impact parameter of tracks relative to the collision vertex in high multiplicity heavy ion collisions. The MVTX utilizes the latest generation of MAPS technology to provide precision tracking with high tracking efficiency over a broad momentum range in the high luminosity RHIC environment. These new capabilities will enable precision measurements of open heavy flavor observables, covering an unexplored kinematic regime at RHIC. The physics program, its potential impact, and recent detector development of the MVTX will be discussed.
\end{abstract}


\end{frontmatter}


\section{Introduction}
sPHENIX detector is a state-of-the-art next generation detector under construction at the Relativistic Heavy Ion Collider (RHIC) at Brookhaven National Laboratory. As stated in the 2015 US Nuclear Physcis Long Range Plan, one focus in the hot QCD area is to probe the inner workings of the Quark-Gluon Plasma (QGP) over a range of momentum and length scales \cite{2015fha}. The complementarity of RHIC and the Large Hadron Collider (LHC) is essential to this goal, and the sPHENIX detector is designed to fulfill this goal. 

The sPHENIX collaboration was formed at the end of 2015, and currently comprises 80 institutions. In 2019, the sPHENIX was granted Project Decision 2/3 (PD-2/3), which approves the start of construction. Also in this year, the sPHENIX became a CERN recognized experiment. The construction and full installation of the detector at RHIC will finish in 2022, and the first data taking will begin in 2023. 

The key physics programs planned at sPHENIX include jet measurements, upsilon spectroscopy, and open heavy flavor measurements. The heavy flavor program relies heavily on the precision vertexing capability provided by the Monolithic Active Pixel Vertex (MVTX) detector. In this presentation, the open heavy flavor physics program at sPHENIX and the MVTX detector developments will be discussed.

\section{sPHENIX Detector}
\begin{figure}
	\centering
	\includegraphics[scale=0.35]{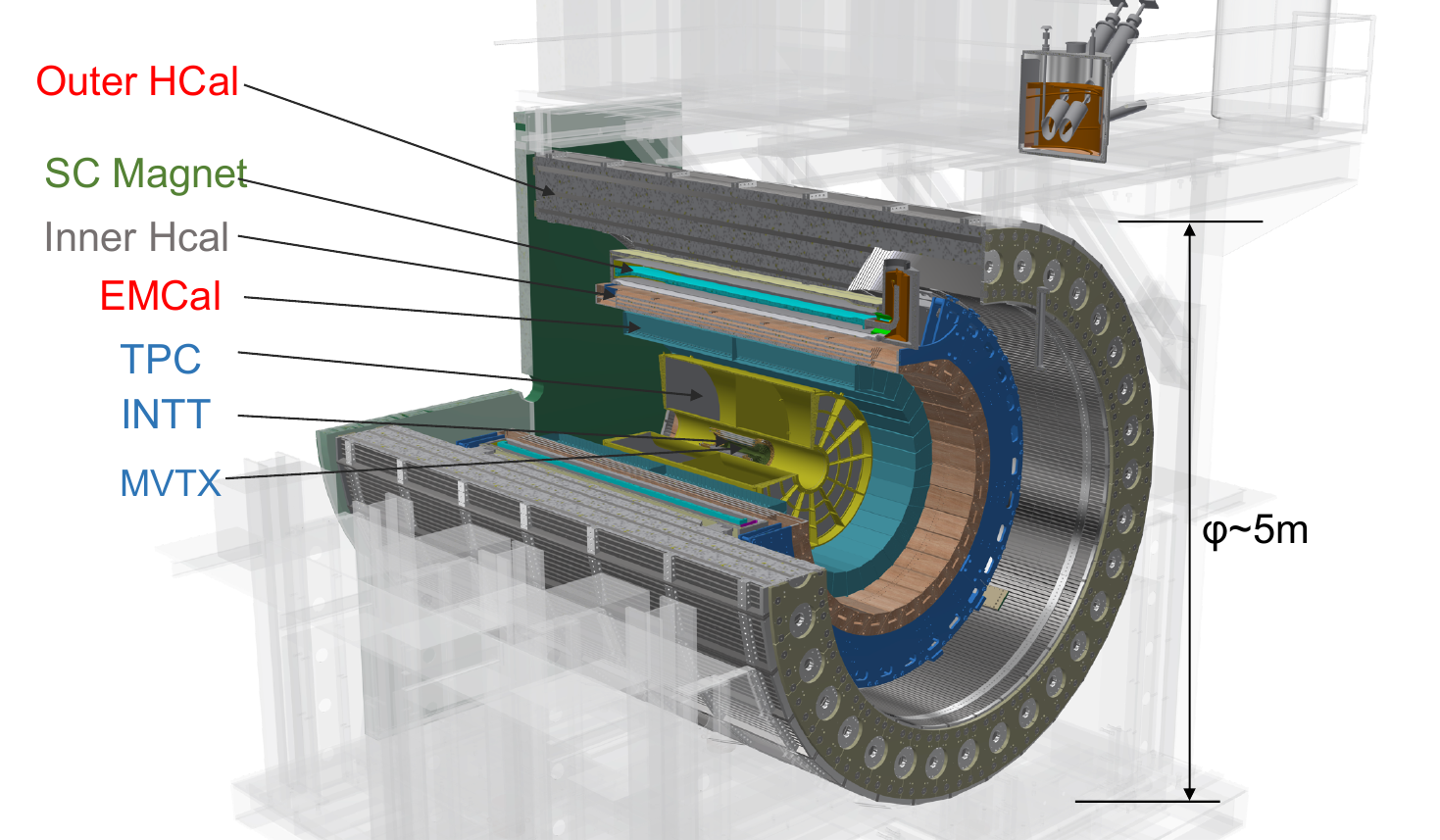}
	\caption{A schematic view of the sPHENIX detector}
	\label{fig:det}
\end{figure}
The sPHENIX detector will be located at RHIC IP-8, where the PHENIX detector was located in the RHIC ring. The detector is designed utilizing proven and cost-effective technology \cite{CDR,TDR}. The schematic layout of the detector is shown in Fig. \ref{fig:det}. The tracking system, from innermost to outer, consists of the MVTX, the Silicon Strip Intermediate Tracker (INTT) and the Time Projection Chamber (TPC). The tracking system will be capable of high precision momentum and displaced vertex measurement. The calorimeter system includes an electromagnetic calorimeter (EMCal), and inner and outer hadronic calorimeters (HCal). Both the tracking and calorimeter systems will cover $|\eta|<1.1$ in pseudorapidity within $|z_{vertex}|<10$ cm with full azimuthal acceptance. The sPHENIX detector trigger rate can record 15 kHz in A+A collisions, and the DAQ data rate is higher than 10 GB/s, so as to fully utilize the increased luminosity of RHIC in the future. Over the planned 3-year operation of sPHENIX, we expect to record 143 billion minimum-bias Au+Au collisions, with comparable reference samples obtained from p+p and p+Au collisions. Studies of the exciting physics potential of even greater statistics have also been performed.

\section{MVTX detector}
\begin{figure}
	\centering
	\includegraphics[scale=0.19]{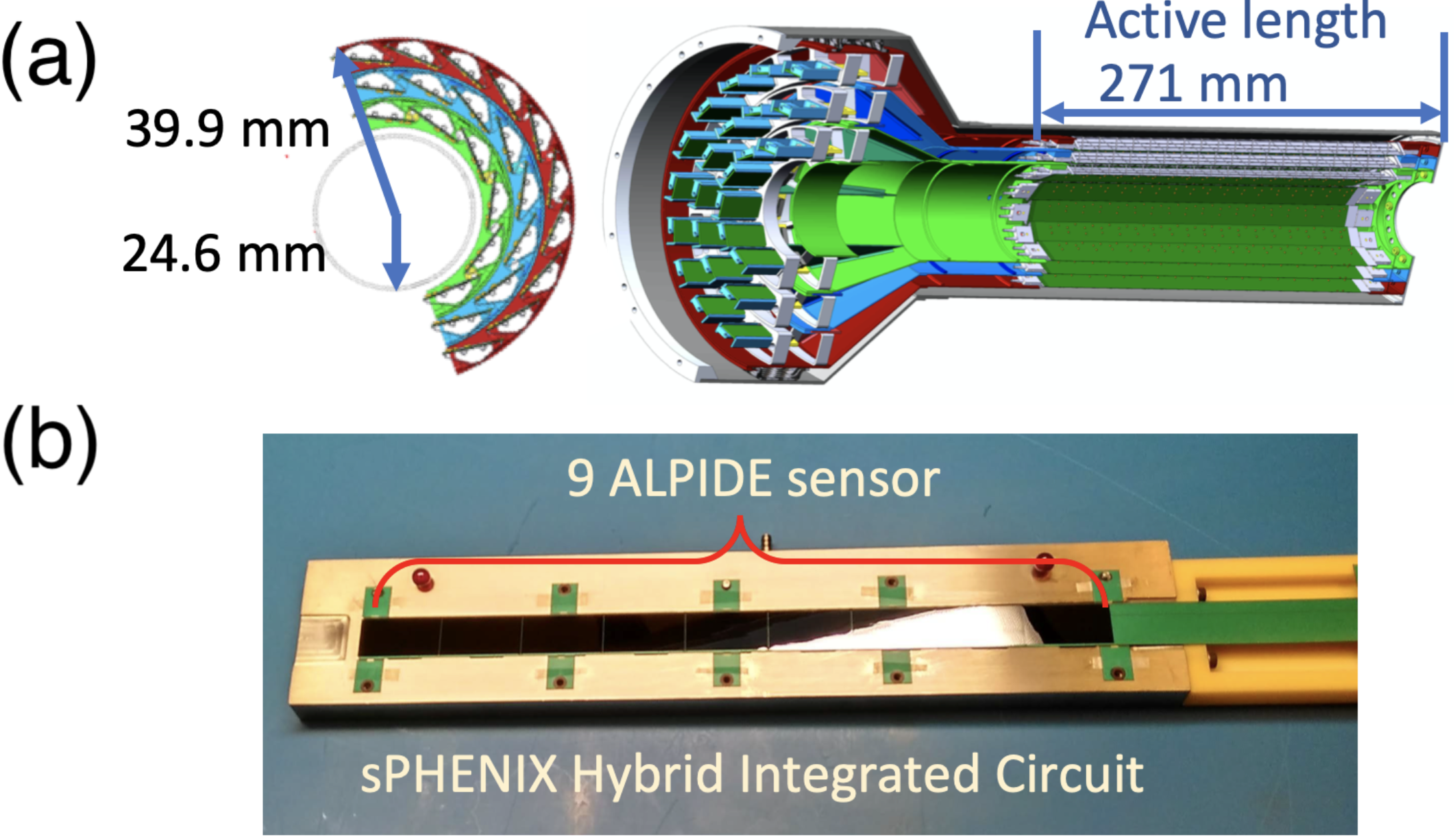}
	\includegraphics[scale=0.16]{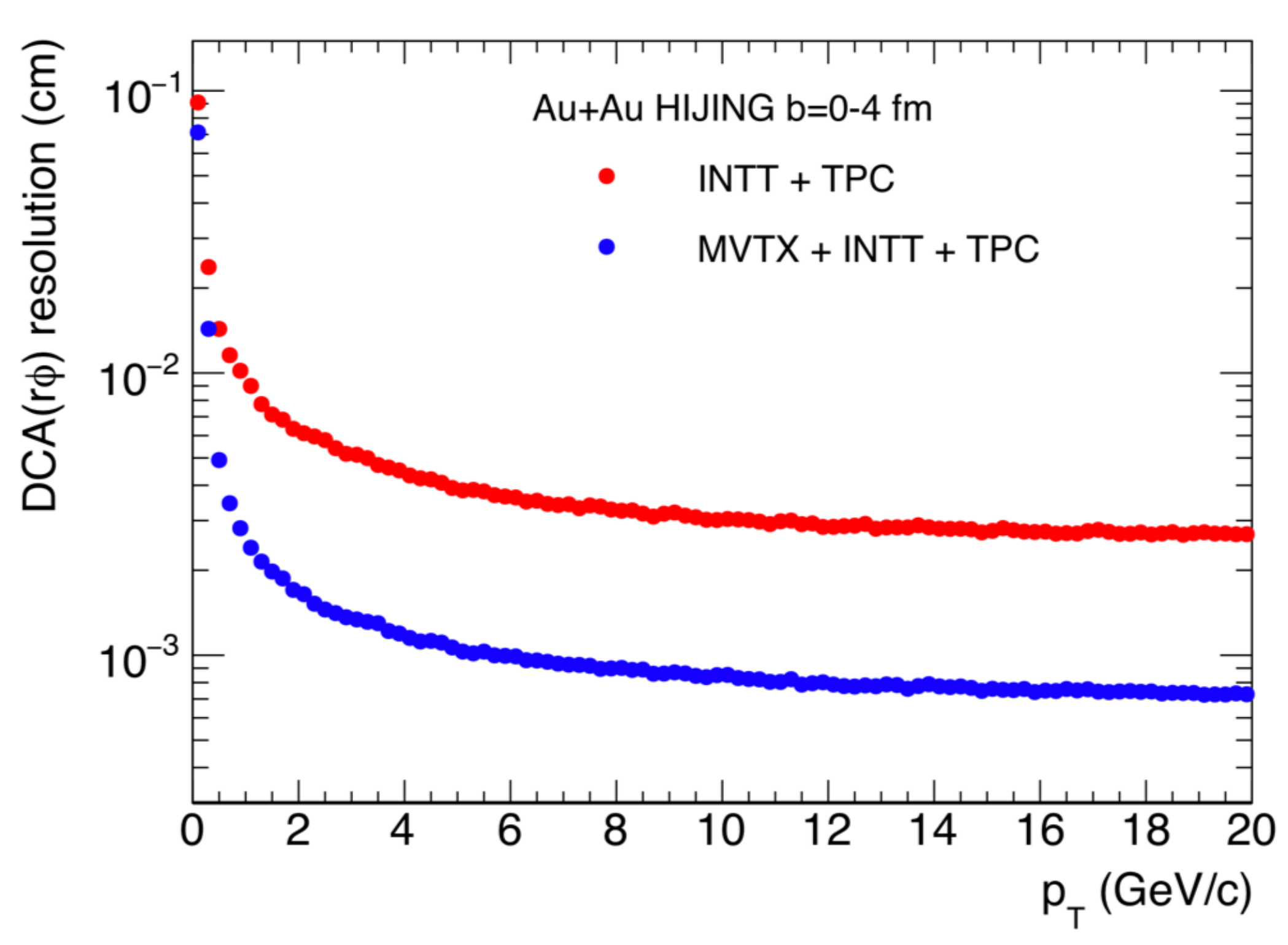}
	\caption{Left: (a) Schematic view  of the  MVTX detector (b) ALPIDE sensor with sPHENIX readout citcuit; Right: Simulated DCA resolution with (blue) and without (red) MVTX added into the tracking system.}
	\label{fig:dcares}
\end{figure}
The high precision vertexing capability of MVTX in combination with the large data sample enables detailed study of many rare heavy flavor observables for the first time at RHIC, such as $B$-hadrons and $b$-jets \cite{MVTXpro}. MVTX is a 3-layer silicon pixel detector based on the ALPIDE sensor, which is a 2nd generation MAPS sensor developed for the ALICE ITS upgrade \cite{alpide}. The MVTX mechanical design is modified from the ITS and the additional readout electronics are developed to fit the sPHENIX envelope. The MVTX schematic view is shown in Fig. \ref{fig:dcares} (left). Full chain beam test, including a complete readout system, had been carried out successfully during May 2019, which confirmed the performance of the system. The full simulation of the sPHENIX detector shows the distance of closet approach (DCA) resolution would be significantly improved with MVTX added into the tracking system, shown in Fig. \ref{fig:dcares} (right). 

\section{Heavy flavor physics program}
Heavy quarks, in particular, the $b$-quark, is a unique hard probe of the QGP as its mass is much higher than the scale of QGP temperature and QCD scales. They are dominantly produced in the initial hard scattering process and experience the whole evolution of the system. Thus heavy-flavor hadrons are regarded as penetrating probes of the QGP. The recent studies on the heavy flavor measurement enabled by sPHENIX detector will be discussed in this section. 
\begin{figure}
	\centering
	\includegraphics[scale=0.3]{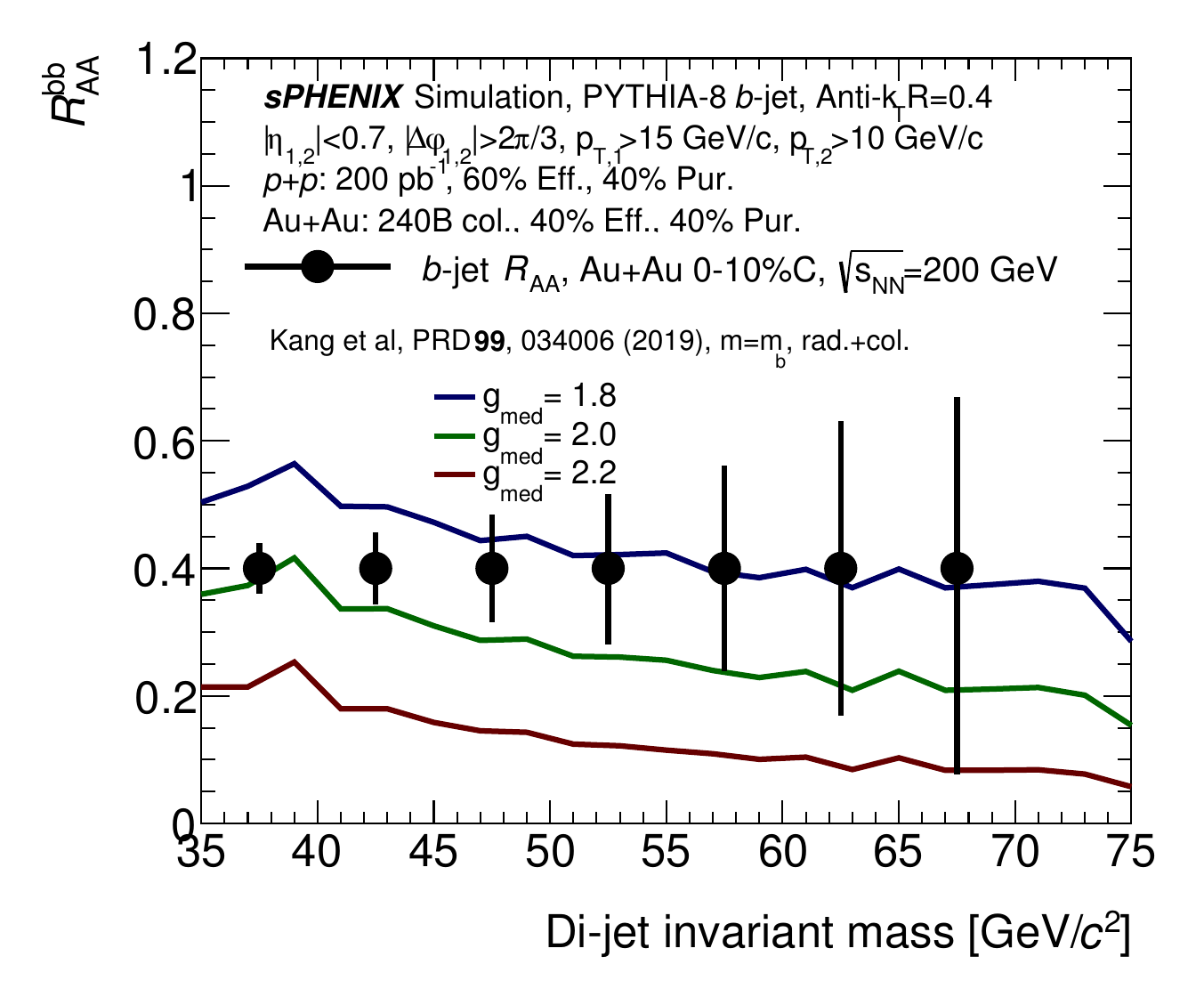}
	\includegraphics[scale=0.3]{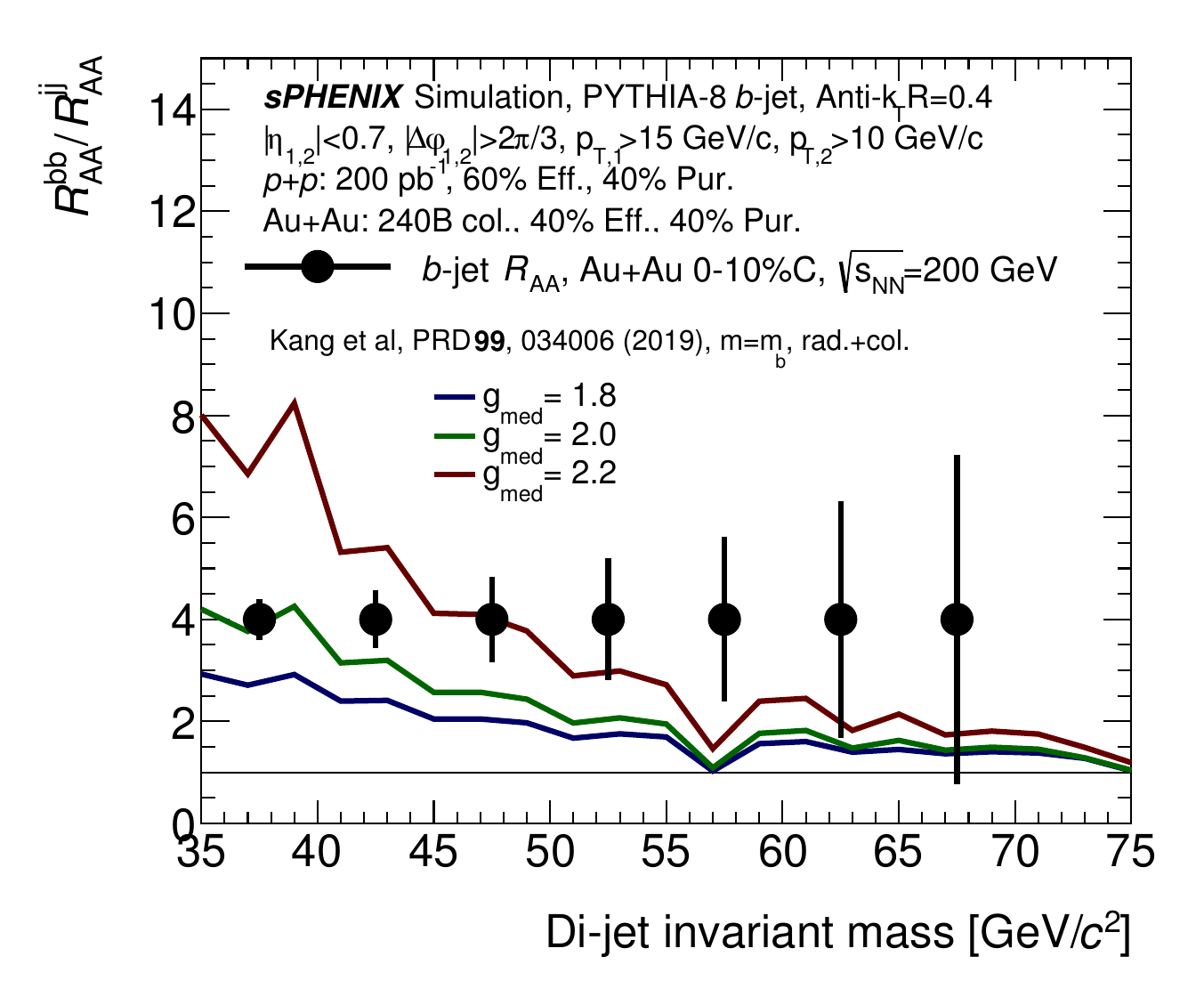}
	\caption{The projected $b$-dijet $R_{AA}$ (left) and the $b$-dijet to inclusive dijet $R_{AA}$ ratio (right) as function as dijet invariant mass in central Au+Au events from the full simulation of the sPHENIX detector. The red, green and blue curves show the theory predictions under different coupling strength between the jet and the medium ($g_{med}$) \cite{KangDijet}.}.
	\label{fig:dijet}
\end{figure}
\begin{figure}
	\centering
	\includegraphics[scale=0.35]{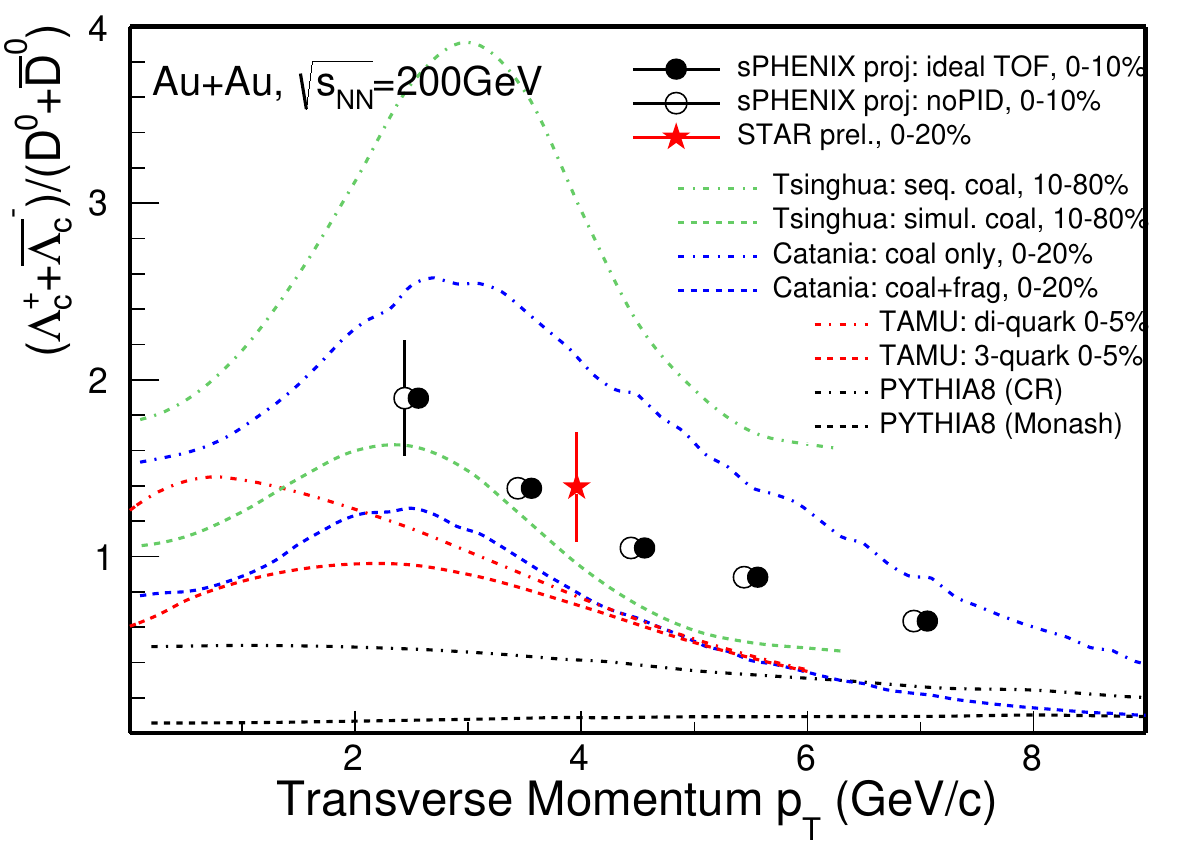}
	\includegraphics[scale=0.22]{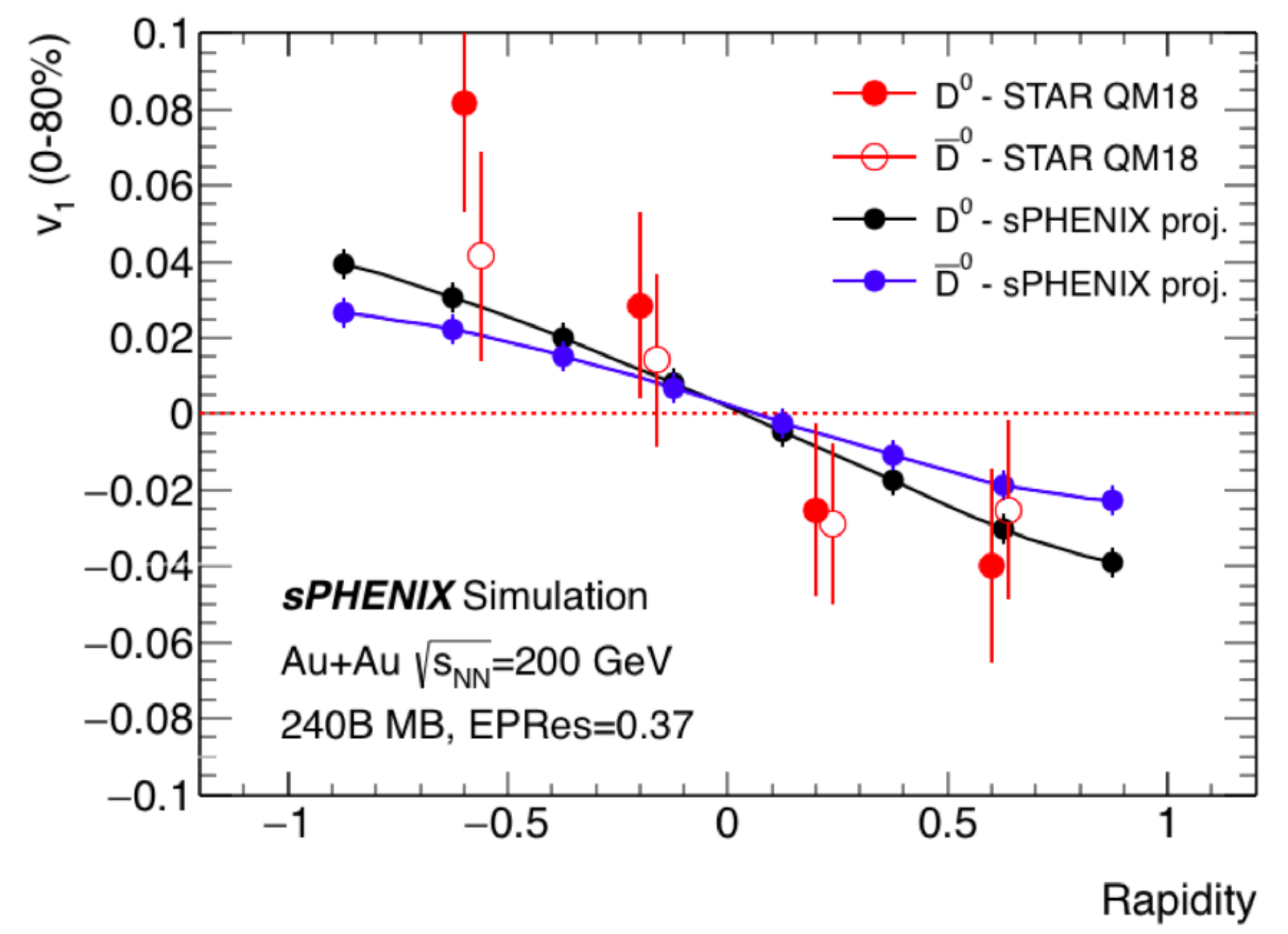}
	\caption{Left: projected $\Lambda_{c}/D^{0}$ in central Au+Au events; Right: projected $D^{0}$ and $\overline{D^{0}}$ $v_{1}$ in central Au+Au events.}
	\label{fig:Lc2D}
\end{figure}
\subsection{Energy loss in QGP medium and QGP transport coefficients}
As shown in Ref. \cite{MVTXpro}, the nuclear modification factor of the bottom quark will be measured over a broad kinematics range from non-perturbative to perturbative regions. Those include non-prompt $D^0$ mesons from $B$-hadron decays at lower $p_T$ region and $b$-jets in $15<p_T<35$~GeV$/c$. Together, they will provide stringent constraint over energy loss models for $b$-quark in the QGP. Recently, Kang and his collaborators further proposed a new $b$-jet observable, the $b$-dijet invariant mass $R^{bb}_{AA}$ \cite{KangDijet}, which shows enhanced sensitivity to transport property and parton-QGP coupling. Furthermore, if we take the ratio of $b$-dijet to inclusive dijet invariant mass $R_{AA}$, the mass effect will be enhanced. The projected $b$-dijet $R^{bb}_{AA}$ and heavy to light flavor dijet $R_{AA}$ ratio at sPHENIX are shown at Fig. \ref{fig:dijet}.

Elliptic flow $v_{2}$ measurements for $B$ mesons would be also be enabled, which will provide clean access to spatial diffusion coefficient at RHIC energy. The projected physics plots for $B$ mesons and inclusive $b$-jets can be found in the MVTX proposal Chapter 2 \cite{MVTXpro}.

\subsection{Heavy flavor hadronization}
Heavy flavor baryon measurements provide us a unique opportunity to understand heavy quark hadronization mechanism in QGP. Strong enhancement of $\Lambda_{c}/D^{0}$ with respect to PYTHIA calculations has been observed from STAR \cite{starLc}. Models based on coalescence mechanism are close to data, but different models still have large differences at low $p_{T}$. In addition, the $\Lambda_{c}/D^{0}$ ratio from STAR measurement indicates that  $\Lambda_{c}$ contributes sizeably to total charm cross section at low $p_{T}$. Thus the possiblility of precise $\Lambda_c$ measurements at sPHENIX is explored. As shown in Fig. \ref{fig:Lc2D} (left), $\Lambda_{c}/D^{0}$ ratio will be measured over a broad $p_{T}$ range from 2-8 GeV/c even in the 10\% most central Au+Au collisions, which will be very useful in differentiating various hadronization modules as shown by the curves. The lowest $p_{T}$ point will be further improved if a PID capability could be introduced to the sPHENIX detector. 

\subsection{Initial condition}
Strong magnetic fields are believed to be generated in the heavy ion collisions. Charm hadrons provide clean access to the initial magnetic field \cite{D0v1th}. Due to the Lorentz force, this field will give $D^{0}$ and $\overline{D^{0}}$ the same magnitude of directed flow ($v_{1}$) but opposite in sign. Current STAR measurement cannot give a solid conclusion whether the $D^{0}$ $v_{1}$ differs from $\overline{D^{0}}$ \cite{D0v1}, since the slope of the $v_{1}$ difference as function as rapidity, $d\Delta v_{1}/dy$, is away from 0 with a $0.36\sigma$ significance even only considering the statistic uncertainty. The projected D meson $v_{1}$ versus rapidity is shown in the Fig. \ref{fig:Lc2D} (right). Assuming the $v_{1}$ difference between $D^{0}$ and $\overline{D^{0}}$ follows the model predictions \cite{D0v1th}, the $d\Delta v_{1}/dy$ will give a non-zero value with a significance of $4.8\sigma$ in the future sPHENIX experiment, which is calculated under a total statistics of 240 billion events and an event plane resolution  of 0.37.

\section{Summary}
To complete the science mission of RHIC and complementary to the LHC measurements in 2020s, the sPHENIX detector will probe the microscope properties of the QGP utilizing hard probes. The sPHENIX detector construction is underway and the first data taking will begin in 2023. The MVTX detector upgrade will open up rich heavy flavor physics opportunities at sPHENIX with many new and high precision observables brought to RHIC energy for the first time. Besides those discussed in this talk, broader topics are also being studied, such as heavy flavor and hadron correlations, $D^{0}$ and $\overline{D^{0}}$ correlation, jet and heavy flavor hadron correlations, heavy flavor tagged jet substructure, total bottom cross-section.

\par
\vspace{3mm}
\noindent $\bold{Acknowledgments}$ The speaker is supported in part by the Natural Science Foundation of China under Grand No. 11675168 and 11720101001, the China Scholarship Council, the U.S. DOE Office of Science.




\bibliographystyle{elsarticle-num}
\bibliography{references}







\end{document}